# Load Restoration Methodology Considering Renewable Energies and Combined Heat and Power Systems

Hayder O. Alwan*, Noor M. Farhan

*University of Baghdad, Iraq*

*\*Corresponding author E-mail: hayderalwan1981@gmail.com*

**Abstract**

Outages and faults cause problems in interconnected power system with huge economic consequences in modern societies. In the power system blackouts, black start resources such as micro combined heat and power (CHP) systems and renewable energies, due to their self-start ability, are one of the solutions to restore power system as quickly as possible. In this paper, we propose a model for power system restoration considering CHP systems and renewable energy sources as being available in blackout states. We define a control variable representing a level of balance between the distance and importance of loads according to the importance and urgency of the affected customer. Dynamic power flow is considered in order to find feasible sequence and combination of loads for load restoration.

## 1. Introduction

The blackout state of a power system is defined as the interruption of electricity generation, transmission, distribution and consumption, when operation of the transmission system or a part thereof is rendered nonfunctional. A power system blackout can cause serious consequences by restricted operation of medical facilities, road, air, and rail traffic congestion, internet breakdown, and interruption in manufacturing processes [1], [2]. Therefore, to reduce the economic impacts and minimize the adverse consequences brought by power system blackout, an efficient power system restoration plan is of utmost importance for power system recovery. For power system restoration, The black-start stage as the first and important stage determines the total ability and speed of power system restoration.

In this stage the power resources which they can start without any need to network power plays an important role. On the other hand, emerging micro grid technology, which enables self-sufficient power systems with distributed energy resources (DERs) such as renewable energies and micro combined heat and power (CHP) systems, provides further opportunities to enhance the self-healing capability [3]. Therefore, renewable energies and CHP systems are potential solutions to improve the ability of system black-start ability.

Increasing number of renewable energy integration to the power system accords a whole host of advantages, such as economic, environmental and technical to power distribution system [4]. These power resources as distributed generation (DG) systems have been studied with different concepts [5]. A versatile energy management algorithm in addition to an efficient forecasting model of renewable energy outputs for these energy resources is necessary to improve their application [7], [8]. However, the other emerging concepts like electric vehicles and demand side management can be considered as an effective part in increasing grid reliability and self-healing ability [9], [10].

Because of the growing environmental awareness and the ratification of the Kyoto Protocol, CHP systems receives again more attention as a way to contribute to the reduction of energy use and emissions. This concept renders a bunch of advantages in case of economic, environmental, and grid reliability [11] considering their optimal scheduling and energy market strategies [12], [13]. A comprehensive study on CHP systems including their output and market operation model is presented in [14]. Authors developed and simulated a novel approach for optimal operation of CHPs in electric network. The feasible operation region they introduced for CHP systems perfectly describes the CHP systems which it could be used for CHP operation modeling for all study areas that they include CHP systems. In this paper we will use this model to indicate CHP output. The next stage after black-start scheduling is network reconfiguration and load restoration.

Significant research has been devoted to solving the load restoration problem [15]–[24]. This problem has also been approached using the meta-heuristic algorithms [15], fuzzy technology [16], multi-agent technologies [17] and other non-structured methods[2]. Authors in [19] proposed a new approach based on optimal power flow technique to minimize customer load shedding aiming to improve load pickup. A software tool to simulate the recovery process of smart grid after an accordance of black out is presented in [20]. In this paper the black-start capability considered as a constant power in the system. Three stage power system restoration considering renewable energies is proposed in [24]. Authors considered renewable energies as negative/positive spinning reserve power to accelerate the power restoration process, details of which can be found in Refs. [1], [2].

### 1.1 Contributions

A review of the literature shows that to date, numerous studies related to load restoration have focused on the aspects of traditional load recovery. However, by increasing renewable energy integrations and CHP systems, impacts of these energy resources on power system restoration should be considered as an available power source which they can play a direct role in power system restoration. Therefore, in modern power system, load restoration considering the renewable energies and CHP systems as an available potential to help for black start stage, need to be carried out further. In this paper, by modeling renewable energy and CHP systems operation, we propose a new system restoration approach. The main contributions of this paper can be listed as: a) Introducing an optimization model for load restoration incorporating renewable energies and CHP systems; b) Considering both load importance and the distance of available black-start resources from load as control variables for optimization problem; and c) Using Greedy algorithm combined with dynamic power flow approach to consider frequency factor and optimize system restoration.

## 2. Problem Formulation

The system restoration process should be formulated mathematically as a system restoration multi-objective optimization problem considering black-start resources operation model such as CHP systems.
The CHP units are assumed to utilize binary gas turbine-water vapor cycles.

$$H_{i,t} + P_{i,t} = z_{i,t} \quad i \in N, t \in T \quad (1)$$

These margins determine their operation points in the system. Therefore, electric output for CHP systems are limited based on their feasible operation region and it should be considered in their operation scheduling.

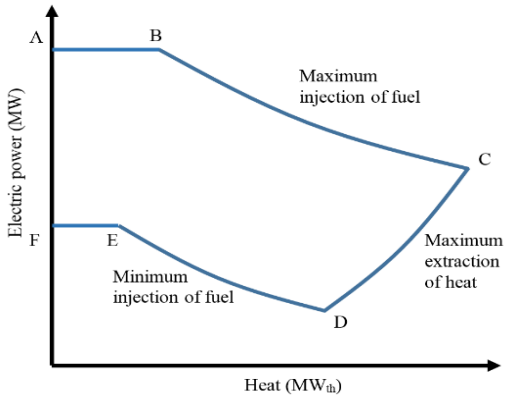

**Fig. 1:** CHP system feasible operation region.

### 2.1 Objective Function

In the power system with renewable energies, loads especially the important loads such as hospitals and communication systems are slated for rapid recovery utilizing self-start capacity afforded by renewable energy sources. Because possibility of restoring power to near customers is higher than distance customers, considering distance of loads from black start resources may provide more accurate and efficient solution for load restoration in distributed system such as the case with renewables. It is rational to consider both important and distance of loads to have more accurate and versatile load restoration plan. Therefore, we propose an optimization model considering load types and their importance and distance between loads and black start resources to maximize the degree of load restoration. The objective function can be formulated as follows:

$$\min f = \sum_{i=1}^{N}[\omega \alpha_i C_i + (1-\omega)C_i/d_i]x_i \quad (2)$$

Where N is the number of loads without service in blackout. $\alpha_i$ Is the importance of $i$th load, $\alpha_i \in [0,1]$. A higher value of $\alpha_i$ represents a higher degree of importance of the particular load. Table I lists the load importance generally is used for load restoration studies [25]. The term $C_i$ is the capacity of $i$th load. $x_i$ Denotes the connection status of the load to the power grid, 0 for no, 1 for yes. $d_i$ is the distance between the $i$th load and the black start resources, which is represented by the number of buses between load and black start resources. $\omega$ is the weight coefficient which denotes the preference of the decision maker in regard to the load importance and the distance between the load and black start resources, $\omega \in [0,1]$. When $\omega =1$, only the effects of black start resources is considered, and $\omega =0$, only the effects of the load importance is considered during the analysis. The term $\omega \alpha_i C_i + (1-\omega)C_i/d_i$ is called as compound load in this paper. This term ties two objectives of maximum load importance and minimum load distance for load recovery, $\alpha_i C_i$ presents the load importance and make sure that the load with high importance is given priority in power restoration. In addition, $C_i/d_i$ models the distance of the load to the black start resources and ensures that the load with shorter distance is prior for load restoration. Based on the operator desire, load restoration could be run to find the best solution focusing on important loads or distance of loads or both of them.

**Table I:** Load importance

| Load type | Importance (*w*) |
|---|---|
| facilities | 1.0 |
| Recovering generators for power system restoration | 0.9 |
| Special loads | 0.6 |
| Supermarkets | 0.43 |
| Counties | 0.12 |

### 2.2 Constraints

It is worthwhile to note that in the power restoration problem, there is also frequency deviation due to unbalance power between generation and load. Therefore, we modeled the power flow equations considering frequency characteristics and other inequality constraints as follows:

$$\Delta P_i = P_{Gi}[1 - K_{Gi}\Delta f] - P_{Li}[1 - K_{Lpi}] = V_i \sum_{j=1}^{N} V_j (G_{ij}cos\theta_{ij} + B_{ij}sin\theta_{ij}) = 0 \quad (3)$$

$$\Delta Q_i = Q_{Gi} - Q_{Li}[1 - K_{Lqi}\Delta f] = V_i \sum_{j=1}^{N} V_j (G_{ij}sin\theta_{ij} + B_{ij}cos\theta_{ij}) = 0 \quad (4)$$

$$|\Delta f| < 1^{Hz} \quad (5)$$

$$|\Delta V_i| < 0.05 \quad i = 1 \dots N \quad (6)$$

$$P_G^{min} < P_{Gi} < P_G^{max} \quad i = 1 \dots N \quad (7)$$

$$Q_G^{min} < Q_{Gi} < Q_G^{max} \quad i = 1 \dots N \quad (8)$$

In power system, frequency control includes three stages: primary, secondary and tertiary frequency control. For the primary control, the steady state frequency error will be calculated and exist due to the proportional control characteristic of system. It should be mentioned that in interconnected power system, automatic generation control (AGC) is implemented to adjust the reference value of the generators output. Since the load pick up will have the direct impact on voltage and frequency, the above mentioned constraints of maximal frequency and voltage deviations which they define by power system operators determine the maximal load that can be restored at one step. The frequency behavior after load restoration is depicted in Fig. 2. In this paper we modeled the frequency control system using dynamic power flow equations.

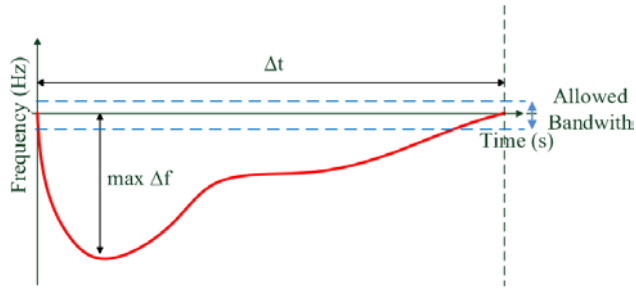

**Fig. 2:** Frequency behavior in load pick up of power system restoration

### 2.3 Dynamic Power Flow with Frequency Deviation

As we mentioned, during restoration the generated power from the recovered generation units is not equal to the load demand. This unbalanced power caused by the difference between generation and demand causes a change in the system frequency. This frequency deviation should consider in dynamic power flow. The diagram of dynamic power flow is presented in Fig. 3.

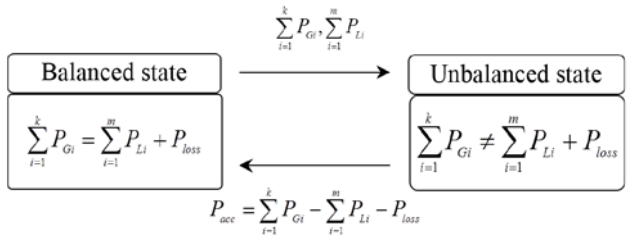

**Fig. 3:** Diagram of dynamic power flow

Where the $P_{acc}$ acceleration power, $P_{Gi}$, restored load $P_{Li}$, line losses $P_{loss}$.

$$P_{acc} = \sum_{i=1}^{k} P_{Gi} - \sum_{j=1}^{m} P_{Li} - P_{loss} \quad (9)$$

$$\Delta f = P_{acc} / (\sum_{i=1}^{k} K_{Gi} + \sum_{j=1}^{m} K_{Li}) \quad (10)$$

Using frequency deviation we will formulate power flow equations to consider unbalance power between generation and load.

## 3. Load Restoration Algorithm

In the proposed algorithm for power system restoration with black start resources, we treated the problem as knapsack problem. In the proposed model, total recovered generation capacity acts as the knapsack capacity, which cannot be exceeded by total amount of recovered loads. Furthermore, the load importance and the distance between the load and black start resources are represented by the weights. In this paper, the Greedy algorithm is applied to solve the load restoration problem, which it is treated as knapsack problem. The Greedy algorithm has been developed for a large number of problems in combinatorial optimization. The Greedy algorithm is typically very easy to describe and code and it has very fast running times which it will be useful in black out restorations as matter of time efficiency. However, in most cases it does not find the global optimum. In load restoration, since the time is more important to system restoration, it is rational to consider a fast algorithm with local optimal solution. The main algorithm of proposed solution is described in following flow diagram.

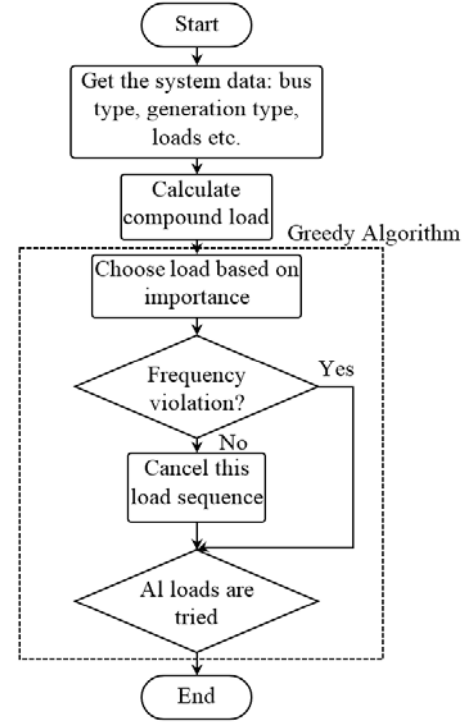

**Fig. 4:** Flowchart of proposed algorithm to power system restoration

## 4. Simulation and Results

To examine proposed method, a part of IEE New England 39 bus bulk system is considered (Fig. 5). This part of system includes 2 thermal units (Gen1 and Gen2) and we considered one CHP system (Gen4) and one wind farm as renewable energy system (Gen 3) to study power system restoration process with CHP and renewable energy systems. The total load is assumed to be about 215.9 MW and 395.7 MVAr. We assumed that the startup time of thermal units in this case is about 2 hours (including energizing process of excitation and ancillary service, etc.). However, the time for switching and energizing the buses and pickup loads is about 2 to 5 minutes which is includes the risk of failures. In the assumed system, loads have been distributed to the load buses assuming their importance factor presented in Table II. We tried to distribute the load importance in the system to study the effect of black-start resources more clearly. In order to simplify the simulation process, we assumed $\pm 10\%$ for the maximum load fluctuation for restored load. In addition, the output for wind farm is modeled based on the 5 min forecasting model provided in [27]. CHP feasible operation regain is used by the example utilized in [14]. And finally, the load model parameters generated randomly considering the acceptable range for their model [24].

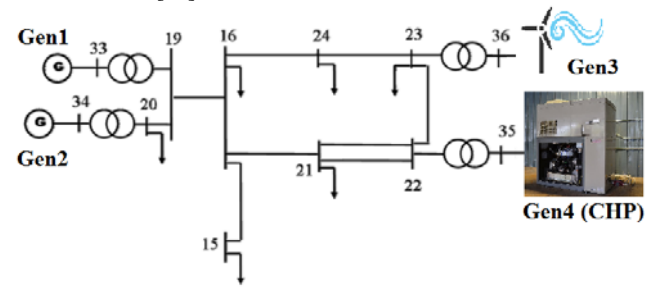

**Fig. 5:** A part of IEEE New England 39 bus power system

**Table II:** Assumed load importance factors

| $\alpha_{15}$ | $\alpha_{16}$ | $\alpha_{20}$ | $\alpha_{21}$ | $\alpha_{23}$ | $\alpha_{24}$ |
|---|---|---|---|---|---|
| 0.6 | 1.0 | 0.4 | 0.7 | 0.2 | 0.3 |

Table III shows the simulation results for the part of IEEE New England system for both load importance objective and load distance objective. Is can be seen for both situations the first step is to schedule the black-start resources to pick up the loads and help to start of non-black start generations. Therefore Gen3 and Gen4 are started for both objectives. Considering load importance, first the bus 22 energizes and then the load at bus 16 the most important bus with importance factor 1 will pick up. And then the load in bus 21 will be energized and picked up with importance factor 0.7. The next step is to energize bus 19 and provide primary start power for Gen1 since the generation of renewable energies and CHP system is not enough to pick up the left over loads. It should be mentioned that the total estimated power system restoration duration is about 4 hours in this procedure. Providing enough energy to start nonblack-start units is the critical step. For the thermal units, after black out if they can operate in islanding mode with their auxiliaries, it is possible to restart and take into the system as hot restart operation which it usually takes about 30 minutes. However, if the restoration process for these units takes long time, they should restart in cold restart operation which usually required about 3 to 24 hours to connect to system to provide electric power. After startup of GEN1, loads in bus 15 and 20 will be picked up and due to the limitation of Gen1 on output power, Gen2 will be start to feed loads on bus 24 and 23. On the other hand, for the power system restoration with objective of load distance, the first load to pick up based on the distance will be the load in bus 23. The interesting fact that we can see from Table III is that the total energy which is provided by wind farm and CHP system is enough for three loads at buses 23, 24, and 21. Therefore, after starting the Gen1 and load pick up at buses 20 and 16, the Gen2 will be start to feed the only load at bus 15.

Table III: Power system restoration actions with load importance and load distance scenario

|     | Actions (Load important) | Actions (Load distance) |
| --- | --- | --- |
| #1 | Black start Gen3 (B36) and Gen4 (B35) | Black start Gen3 (B36) and Gen4 (B35) |
| #2 | Energize(B22) | Load pickup (B23) |
| #3 | Load pickup (B16) | Load pickup (B24) |
| #4 | Load pickup (B21) | Energize(B22) |
| #5 | Energize(B19) | Load pickup (B21) |
| #6 | Start Gen1 | Energize(B19) |
| #7 | Load pickup (B15) | Start Gen1 |
| #8 | Load pickup (B20) | Load pickup (B20) |
| #9 | Start Gen2 | Load pickup (B16) |
| #10 | Load pickup (B24) | Start Gen2 |
| #11 | Load pickup (B23) | Load pickup (B15) |

Table IV shows the simulation results for different control variable of ω to study the effect of different objectives of load importance and load distance on power system restoration sequence. It is clear that with increasing the value of control variable ω meaning to increase the weighting factor of load importance, the load restoration sequence restructures based on the importance factor of loads. However, in some ω, such as ω=0.1 and ω=0.2, there is not a significance difference in load restoration sequence. This is due to fact that for these sequences we should consider a big change based on load distance or load importance to change in load restoration sequence. It should be mentioned that in power system restoration it is possible to have a generation potential less than the load. In this situation, in load pick up we may have a big frequency change in the system and it may violate to frequency constraint. Therefore we will not be able to feed those loads. For instance, in power restoration with ω=0.0, it Gen2 could not restart and connect to the network, will not be able to pick up load at bus 15, which has the load importance equal to 0.6.

Table IV: Power restoration sequence actions with different control variable ω

| Control variable ω | Restoration Actions |
| --- | --- |
| 0.0 | B36, B35, B23, B24, B22, B21, B19, B33, B20, B15, B34, B16 |
| 0.1 | B36, B35, B23, B24, B22, B21, B19, B33, B20, B34, B16, B15 |
| 0.2 | B36, B35, B23, B24, B22, B21, B19, B33, B20, B34, B16, B15 |
| 0.3 | B36, B35, B23, B24, B22, B21, B19, B33, B16, B34, B20, B15 |
| 0.4 | B36, B35, B23, B24, B22, B21, B19, B33, B16, B34, B20, B15 |
| 0.5 | B36, B35, B22, B23, B22, B16, B19, B33, B21, B34, B20, B15 |
| 0.6 | B36, B35, B22, B21, B19, B16, B33, B20, B15, B34, B23, B24 |
| 0.7 | B36, B35, B22, B21, B19, B16, B33, B20, B15, B34, B23, B24 |
| 0.8 | B36, B35, B22, B21, B16, B19, B33, B15, B20, B34, B24, B23 |
| 0.9 | B36, B35, B22, B21, B16, B19, B33, B15, B20, B34, B24, B23 |
| 1.0 | B36, B35, B22, B16, B21, B19, B33, B15, B20, B34, B24, B23 |

It is worthy to note that although the load at bus 16 has higher importance in comparison with other load, in most of the cases load at bus 15 is picked up as the first load in load restoration. This is due to the distance of the bus 16 from black start resources. In fact, the load in the energized buses is selected to be restored by considering the load priority and available power of online generators.

## 5. Conclusion

In this paper a power system restoration with renewable energies and CHP systems as black start resources presented. We considered a multi objective process with the load importance and distance of loads simultaneously. We also used a control variable to adjust the balance between considered objectives for power system restoration. Defining the compound load as fitness function provides a precise measure of both the load importance and distance of loads from black start resources which can be used by power system planers to adjust scheduling process based on their preferences. Using the proposed load restoration algorithm, for each control variable ω and certain available power generation, the suitable load sequences for load restoration can be obtained. For proof of the concept, we just considered a part of IEEE New England network with black start resources. Future work may include a detailed model of power system resources as distributed generations. In addition, uncertainty of demand and output of black start resources, especially renewable energies in power system restoration scheduling could be considered. The proposed approach can help power system operators and planners to construct a versatile power system restoration plan and make better decision during the system restoration.

## References

[1] D. Lindenmeyer, H. W. Dommel, and M. M. Adibi, "Power system restoration - a bibliographical survey," *Int. J. Electr. Power Energy Syst.*, vol. 23, no. 3, pp. 219–227, 2001.
[2] Y. Liu, R. Fan, and V. Terzija, "Power system restoration: a literature review from 2006 to 2016," *J. Mod. Power Syst. Clean Energy*, vol. 4, no. 3, pp. 332–341, 2016.
[3] A. Majzoobi and A. Khodaei, "Application of Microgrids in Supporting Distribution Grid Flexibility," *IEEE Trans. Power Syst.*, vol. 32, no. 5, pp. 3660–3669, Sep. 2017.
[4] A. Majzoobi, M. Mahoor, and A. Khodaei, "Microgrid Value of Ramping," in *IEEE International Conference on Smart Grid Communications, Dresden, Germany,* 2017.
[5] A. Jalali, S. K. Mohammadi, H. Sangrody, and A. R. Karlsruhe, "DG-embedded radial distribution system planning using binary-selective

PSO," in *2016 IEEE Innovative Smart Grid Technologies - Asia (ISGT-Asia)*, 2016, pp. 996–1001.

[6] M. Mahoor, N. Iravani, S. M. Salamati, A. Aghabali, and A. Rahimi-Kian, "Smart energy management for a micro-grid with consideration of demand response plans," in *2013 Smart Grid Conference (SGC)*, 2013, pp. 125–130.

[7] M. Alanazi, M. Mahoor, and A. Khodaei, "Two-stage hybrid day-ahead solar forecasting," in *2017 North American Power Symposium (NAPS)*, 2017, pp. 1–6.

[8] M. Mahoor, Z. S. Hosseini, A. Khodaie, and D. Kushner, "Electric Vehicle Battery Swapping Station," in *CIGRE Grid of the Future Symposium, 2017, Cleveland, OH.*, 2017.

[9] M. Mahoor, F. R. Salmasi, and T. A. Najafabadi, "A Hierarchical Smart Street Lighting System With Brute-Force Energy Optimization," *IEEE Sens. J.*, vol. 17, no. 9, pp. 2871–2879, May 2017.

[10] M. De Paepe, P. D'Herdt, and D. Mertens, "Micro-CHP systems for residential applications," *Energy Convers. Manag.*, vol. 47, no. 18–19, pp. 3435–3446, Nov. 2006.

[11] H. Cho, R. Luck, S. D. Eksioglu, and L. M. Chamra, "Cost-optimized real-time operation of CHP systems," *Energy Build.*, vol. 41, no. 4, pp. 445–451, Apr. 2009.

[12] S. Vejdan and S. Grijalva, "The Value of Real-Time Energy Arbitrage with Energy Storage Systems," in *Power and Energy Society General Meeting (PESGM)*, 2018.

[13] H. R. Sadeghian and M. M. Ardehali, "A novel approach for optimal economic dispatch scheduling of integrated combined heat and power systems for maximum economic profit and minimum environmental emissions based on Benders decomposition," *Energy*, vol. 102, pp. 10–23, May 2016.

[14] Hayder O. Alwan and Q. S-Al-Sabbagh, "Detection of Static Air-Gap Eccentricity in Three Phase induction Motor by Using Artificial Neural Network (ANN)," Vol. 7-Issue 5 (May-2017) *Int. Journal of Engineering Research and Application (IJERA)*, ISSN:2248-9622,www.ijera.com

[15] A. Augugliaro, L. Dusonchet, and E. Riva Sanseverino, "Multiobjective service restoration in distribution networks using an evolutionary approach and fuzzy sets," *Int. J. Electr. Power Energy Syst.*, vol. 22, no. 2, pp. 103–110, 2000.

[16] F. Ren, M. Zhang, D. Soetanto, and X. Su, "Conceptual design of a multi-agent system for interconnected power systems restoration," *IEEE Trans. Power Syst.*, vol. 27, no. 2, pp. 732–740, 2012.

[17] T. T. Ha Pham, Y. Bésanger, C. Andrieu, N. Hadjsaid, M. Fontela, and B. Enacheanu, "A new restoration process in power systems with large scale of dispersed generation," *Proc. IEEE Power Eng. Soc. Transm. Distrib. Conf.*, pp. 1185–1190, 2006.

[19] R. Ribeiro *et al.*, "An Advanced Software Tool to Simulate Service Restoration Problems: A case study on Power Distribution Systems," *Procedia Comput. Sci.*, vol. 108, pp. 675–684, 2017.

[20] Hayder O. Alwan, N. M. Farhan, and Q. S-Al-Sabbagh, "Design of Data Aquistion interface circuit used in Detection Inter-turn Faultin Motorbased onMotor Current Signature Analysis (MCSA)Technique," Vol. 7-Issue 6 (June-2017) *Int. Journal of Engineering Research and Application (IJERA), ISSN:2248-9622,www.ijera.com*

[21] Y. Jiang *et al.*, "Blackstart capability planning for power system restoration," *Int. J. Electr. Power Energy Syst.*, vol. 86, pp. 127–137, 2017.

[22] Hayder O. Alwan and Q. S-Al-Sabbagh, "Various Types of Faults and Their Detection Techniques in Three Phase Induction Motors Fault," *Int. Journal of Engineering Research and Application,* Vol. 7-Issue 5 (May-2017) *Int. Journal of Engineering Research and Application (IJERA)*, ISSN:2248-9622,www.ijera.com

[23] C. Shen, P. Kaufmann, C. Hachmann, and M. Braun, "Three-stage power system restoration methodology considering renewable energies," *Int. J. Electr. Power Energy Syst.*, vol. 94, pp. 287–299, 2018.

[24] W. Sun, C.-C. Liu, and L. Zhang, "Optimal Generator Start-Up Strategy for Bulk Power System Restoration," *IEEE Trans. Power Syst.*, vol. 26, no. 3, pp. 1357–1366, Aug. 2011.

[25] D. Rodriguez Medina *et al.*, "Fast Assessment of Frequency Response of Cold Load Pickup in Power System Restoration," *IEEE Trans. Power Syst.*, vol. 31, no. 4, pp. 3249–3256, Jul. 2016.

[26] G. S. Piperagkas, A. G. Anastasiadis, and N. D. Hatziargyriou, "Stochastic PSO-based heat and power dispatch under environmental constraints incorporating CHP and wind power units," *Electr. Power Syst. Res.*, vol. 81, no. 1, pp. 209–218, 2011.